\documentstyle[numreferences,graphicx,editedvolume]{crckapb} 

    \newcommand{\vecb}[1]{\mbox{\bf#1}}
   
   \newcommand{\vecbm}[1]{\mbox{\boldmath#1}}
   
   \newcommand{\mra}  {\rightarrow}

\begin{opening}
\title{ PHASE TRANSITIONS\protect\\ WITHOUT THERMODYNAMIC LIMIT}
\subtitle{The Crucial R\^ole of Possible and Impossible Fluctuations\protect\\
The Treatment of Inhomogeneous Scenaria in the Microcanonical Ensemble}

\author{D.H.E. Gross}
\institute{Hahn-Meitner-Institute Berlin,Bereich 
Theoretische Physik,\protect\\Glienickerstr.100,14109 Berlin, Germany\protect
\\ and\protect\\
Freie Universit\"at Berlin, Fachbereich Physik}

\end{opening}
\runningtitle{PHASE TRANSITIONS WITHOUT THERMODYNAMIC LIMIT}
\begin{document}
\section{Introduction}
{\em ``In the thermodynamic limit the canonical and the microcanonical
ensemble are equivalent in all details and generality. ''} Statements
like this are found in many textbooks of statistical
thermodynamics. It is the purpose of this contribution to show that
this is not so, and, more importantly, that the microcanonical
ensemble allows for significant insight into the mechanism of first
order phase-transitions which is hidden in the canonical
ensemble. E.g.: at the liquid-gas transition under given pressure
large and fluctuating spatial inhomogeneities are created. This
surface entropy $S_{surf}$ governs the surface tension. At the
critical point $T_{cr}S_{surf}$ compensates even the surface energy of
the ground state leading to the vanishing of the surface tension.

In the canonical ensemble at given pressure, however, inhomogeneities
become suppressed by
\begin{equation}
\sim e^{-N^{2/3}f_{surf}/T_{boil}},\label{suppress}
\end{equation}
relative to the pure liquid or pure gas configurations, where $N$ is the number
of particles and $f_{surf}$ the free energy of the interphase surface per
surface atom.( In the following we use small letters for the energy or entropy
per atom.) Consequently, macroscopic interfaces do not exist in the
canonical ensemble.

This has a very practical consequence:{\em``If phase transitions in
the experimental world would be at constant temperature not under
controlled supply of energy''} i.e. energy constraints unimportant as
is often claimed, we would not be able to see a pot with boiling
water.

Other common wisdoms are: {\em ``Phase transitions exist only in the
thermodynamic limit.''} This is certainly true within the canonical
ensemble where phase transitions are indicated by a singularity {\em
point} as functions of the temperature. For finite systems, however,
the canonical partition sum
\begin{equation}
Z(T)=\sum_i{e^{-N  e_i/T}}\label{partS}\approx\int{
e^{N(s_N( e)- e/T)}N~de}.
\end{equation}
is analytic in $\beta=1/T$. In the micro ensemble the transition occurs over a
whole {\em region} in energy, {\em relatively independent of the system size
$N$}.

{\em "S must be additive (extensive).
For microscopic objects like atoms and cosmic like stars thermodynamics is
inappropriate !"\cite{lieb98a}}
In contrast, I will show in the following that phase transitions are
well defined and can be well classified into first order or continuous
transitions for pretty small systems like nuclei or some $100-1000$ atoms in
the {\em microcanonical} ensemble.  Moreover, for the liquid--gas transition in
sodium, potassium, and iron the microcanonical transition temperature $T_{tr}$,
the specific latent heat $q_{lat}$, and the specific interphase surface tension
$\sigma_{surf}$ are for ca. $1000$ atoms close to their bulk values
\cite{gross157}.  Nothing special qualifies the liquid--gas transition in
infinite systems (at least for the above three realistic systems). This is also
so in several toy-models of statistical mechanics: In \cite{gross150} we found
that for a microcanonical 2-dim Potts model with $q=10$ states/lattice-point
$T_{tr},q_{lat},\sigma_{surf}$ are within a few percent close to their bulk
values for some hundred spins.  This astonishing fast scaling of the
microcanonical transition parameters with the particle number towards their
bulk values was explained by H\"uller and Promberger in a recent paper
\cite{promberger96} by the fact that the trivial factor $N$ in the exponent of
the Laplace transform (\ref{partS}) is the main origin of canonical finite
size scaling, which has nothing to do with the physics of the phase
transition: Even if one replaces $s_N(e)$ in (\ref{partS}) by $s_\infty( e)$
of the infinite system one gets
practically the same scaling of the specific heat with $N$.

In macroscopic thermodynamics extensive variables like total energy,
mass, and charge are fixed only in the mean by the intensive $T$,
$\mu_N$, $\mu_Z$ \cite{hill64}. In nuclear physics they are, however,
strictly conserved. Nuclear systems are too small to ignore the
fluctuations of conserved quantities when using the usual macroscopic
thermodynamic relations as Hill suggests even if the fluctuations are
somehow $\sim 1/\sqrt{N}$. Especially, if we are interested in phase
transitions, we must be more careful: At phase transitions the effect
of the fluctuations is usually even much larger, see below. Therefore,
we test here statistical mechanics under extremely {\em different}
conditions than we are used to.  The foundations of thermodynamics
must be revisited, extended and deepened. The constraints due to
global conservation laws must carefully be respected and separated
from the still persistent statistical fluctuations. What is more
remarkable and often overseen: Even in macroscopic systems statistical
fluctuations{\em /particle} do not vanish and are usually large at
phase-transitions of first order.  {\em I.e. even for phase
transitions in the bulk (the standard field of thermodynamics)
fluctuations and their constraints are important.}  Because it takes
fluctuations serious our extended statistical thermodynamics can be
applied to phase transitions in small as well in large
systems. Especially the development of phase transitions with the size
can then be followed from small to the bulk. Microcanonical
thermodynamics even sheds new light on details of the mechanism
leading to phase transitions in the bulk (sect.3). Moreover, in
contrast to \cite{lieb98a} it allows to address thermodynamically
unstable like gravitating systems. There is some chance, this is
possible even without invoking an artificial container.

Another example illuminates what I want to emphasize: Angular momentum is
certainly one of the more exotic conservation laws in thermodynamics. In
reality, however, it is much more important than one might think: In
astrophysical systems like e.g. collapsing cosmic hydrogen clouds it may decide
if a single star or a rotating multiple star system is born.

\section{Microcanonical thermodynamics}

Microcanonical thermodynamics explores the topology of the N-body
phase space and determines how its volume $W(E,N)=e^S$ depends on the
fundamental globally conserved quantities namely total energy
$E=N* e$, angular momentum $\vecb{L}$, mass (number of atoms
$N$), charge $Z$, linear momentum $\vecb{p}$, and last not least, if
necessary, the available spatial volume $V$ of the system.  This definition is
the basic starting point of any statistical thermodynamics since
Boltzmann\cite{boltzmann}. If we do not know more about a complicated
interacting N-body system but the values of its globally conserved macroscopic
parameters, the probability to find it in a special phase space point (N-body
quantum state) is uniform over the accessible phase space. This is an {\em
entirely mechanistic} definition. It is of course a completely separated and
difficult question, outside of thermodynamics, if and how a complicated
interacting many-body system may explore its entire allowed phase space. In a
nuclear or cluster collisions it is not really necessary that every dynamical
path visits the entire possible phase space.  It is sufficient that the
evolution of an {\em ensemble} of many {\em replica}, one after the other, of
the same system under identical macroscopic initial conditions follows the
structure of the underlying N-body phase space. It is ergodic in the same sense
as the dynamics of a falling ball is ergodic on Galton's nailboard. In nuclear
fragmentation the ergodicity is presumably due to the strong and short ranged
friction between moving nuclei in close proximity.  Friction between atomic
clusters is yet unknown but probably it exists there also.\\ {\em 2.1)~~
Statistical ensemble}\\ Before we proceed, we have to emphasize the concept of
the statistical {\em ensemble} under strictly conserved energy, angular
momentum, mass, and charge.  As we explained above, microcanonical
thermodynamics describes the dependence of the volume $e^S$ of the -- at the
given energy $E$, small interval $\delta E$, -- accessible phase space on the
globally conserved energy, mass, and charge.  Each phase space cell of size
$(2\pi\hbar)^{3N-6}$ corresponds to an individual configuration (event) of our
system. We realize the ensemble by replica {\em in time} under identical
macroscopic conditions (events).  This is different from Hill \cite{hill64} who
assumes a macroscopic noninteracting supersystem of many identical copies of
the system under consideration. This would be impossible e.g. for rotating or
gravitating systems.  Clearly, the volume $e^S$ of the phase space is the sum
(ensemble) of all possible phase space cells compatible with the values of
energy etc..

While the conserved, extensive quantities, energy, momentum,
number of particles, and charge can be determined for each individual
configuration of the system, i.e.  at each phase-space {\em point} or
each event, this is not possible for the phase space {\em volume}
$e^S$, i.e. the entropy $S(E,V,N)$ and all its increments like the
temperature $T=(\partial S(E,V,N)/\partial E)^{-1}$, the pressure
$P(E,V,N)=T\partial S(E,V,N)/\partial V$, and the chemical potential
$\mu =-T\partial S(E,V,N) /\partial N$. They
are {\em ensemble averages}. Only in the thermodynamic limit $N\mra\infty$
may e.g. the temperature be
determined in a single configuration by letting the energy flow into a
small thermometer.  For a finite system, e.g.  a finite atomic
cluster, the temperature, its entropy, its pressure can only be
determined as ensemble averages over a large number of individual
events. E.g. in a fusion of two nuclei the excitation
energy in each event is given by the ground-state Q-values plus the incoming
kinetic energy whereas the temperature of the fused compound nucleus
is determined by measuring the kinetic energy {\em spectrum} of decay
products which is an average over many decays.  It is immediately
clear that the size of $S$ is a measure of the {\em fluctuations} of
the system.\\~\\
{\em 2.2)~~ Microcanonical signal for a phase transition: 
The caloric curve.}\\
The most dramatic phenomena in thermodynamics are phase transitions. I will
try to interpret them microcanonically as peculiarities of the topology of the
N-body phase space. I will avoid the concept of the thermodynamic limit as I
believe that this is not really essential to understand phase
transitions.  We will see that details about the transitions become more
transparent in finite systems.  Then however, one needs a modified definition
of phase transitions.

In \cite{gross72,gross154} we introduced a new criterion of phase transitions,
which avoids any reference to the thermodynamic limit and can also be used for
finite systems: The anomaly of the microcanonical caloric equation of state
$T(E/N)$ where $\partial T/\partial E \le 0$ i.e.  where the familiar monotone
rise of the temperature with energy is interrupted. {\em Rising the energy
leads here to a cooling of the system}.  This anomaly corresponds to a convex
intruder in $S(E)=\int{1/T(E)dE}$. At energies where $S(E)$ is convex the
system would spontaneously divide into two parts and gain
entropy eg.: $(S(E_1)+S(E_2))/2>S((E_1+E_2)/2)$, iff the creation of the
interface would not {\em cost} an extra entropy $\Delta S_{surf}$.  Because
$\Delta s_{surf}$ per atom vanishes $\propto N^{-1/3}$ in the thermodynamic
limit $s_\infty(e)$ is concave as demanded by van Hove's theorem \cite{vanhove49} and
by the second law of thermodynamics.

Very early the anomaly of the caloric curve $T(E/N)$ was taken as signal for a
phase transition in small systems in the statistical theory of
multi-fragmentation of hot nuclei by Gross and collaborators
\cite{gross72,gross75} and the review article \cite{gross95}.  Bixton and
Jortner \cite{bixton89} linked the back-bending of the microcanonical caloric
curve to strong bunching in the quantum level structure of the many-body system
i.e. a rapid and sudden opening of new phase space when the energy rises.
Their paper offers an interesting analytical investigation of this connection.

A phase transition of first order is characterized by a sine-like oscillation,
a ``back-bending'' of $T( e=E/N)$ c.f. fig.1. As shown below, the
Maxwell-line which divides the oscillation of $\partial S/\partial
E=\beta( e)=1/T$ into two opposite areas of equal size gives the
inverse of the transition temperature $T_{tr}$, its length the specific latent
heat $q_{lat}$, and the shaded area under each of the oscillations is the loss
of specific entropy $\Delta s_{surf}$ as mentioned above. The latter is
connected to the creation of macroscopic interphase surfaces, which divide
mixed configurations into separated pieces of different phases, e.g.  liquid
droplets in the gas or gas bubbles in the liquid.  Even nested situations are
found like liquid droplets inside of crystallized pieces which themselves are
swimming in the liquid in the case of the solid -- liquid transition, see e.g.
the experiments reported in \cite{grimsditch96}. {\em I.e. at phase transitions
of first order inhomogeneous ``macroscopic or collective'' density fluctuations
are common,} boiling water is certainly the best known example. Phase-dividing
surfaces of macroscopic size exist where many atoms collectively constitute a
boundary between two phases which cause the reduction of entropy by $\Delta
s_{surf}$.\\~\\ 
{\em 2.3)~~ "Maxwell" construction of $T_{tr}$, $q_{lat}$, $\sigma_{surf}$}\\
As the entropy is the integral of $\beta( e)$:
$s( e)=\int_0^{ e}{\beta{( e')}d e'}$ it is
a concave function of $ e$ ($\partial^2s/\partial e^2
=\partial\beta/\partial e < 0$) as long as $T( e)=\beta^{-1}$
shows the usual monotonic rise with energy. In the pathological back-bending
region of $\beta( e)$ the entropy $s( e)$ has a convex
intruder of depth $\Delta s_{surf}$ \cite{gross150}. At the beginning
($\ge e_1$) (c.f. fig.1c) of the intruder the specific entropy
$s( e)$ is reduced compared to its concave hull, which is the double
tangent to $s( e)$ in the points $ e_1$ and $ e_3$.
The derivative of the hull to $s( e)$ follows the Maxwell-line in the
interval $ e_1\le e\le e_3$.  In the middle,
($ e_2$), when the separation of the phases is fully established this
reduction is maximal $=\Delta s_{surf}$ and at the end of the transition
($ e_3$) when the intra-phase surface(s) disappears $\Delta s_{surf}$
is gained back.  Consequently, the two equal areas in $\beta( e)$ are
the initial loss of surface entropy $\Delta s_{surf}$ and the later regain of
it.  Due to van Hove's theorem  this convex intruder of $s( e)$ must
disappear in the thermodynamic limit which it will do if $\Delta s_{surf} \sim
N^{-1/3}$. This is why a transition of first order may easier be identified in
finite systems where the intruder can still be seen.  The intra-phase surface
tension is related to $\Delta s_{surf}$ by $\gamma_{surf}=\Delta
s_{surf}*N*T_{tr}/\mbox{surf.-area}$. 
\cite{gross153}\}.\\
\begin{minipage}[t]{6.5cm}
\includegraphics*[bb = 7 168 480 514, angle=-90, width=6.5cm,  
clip=true]{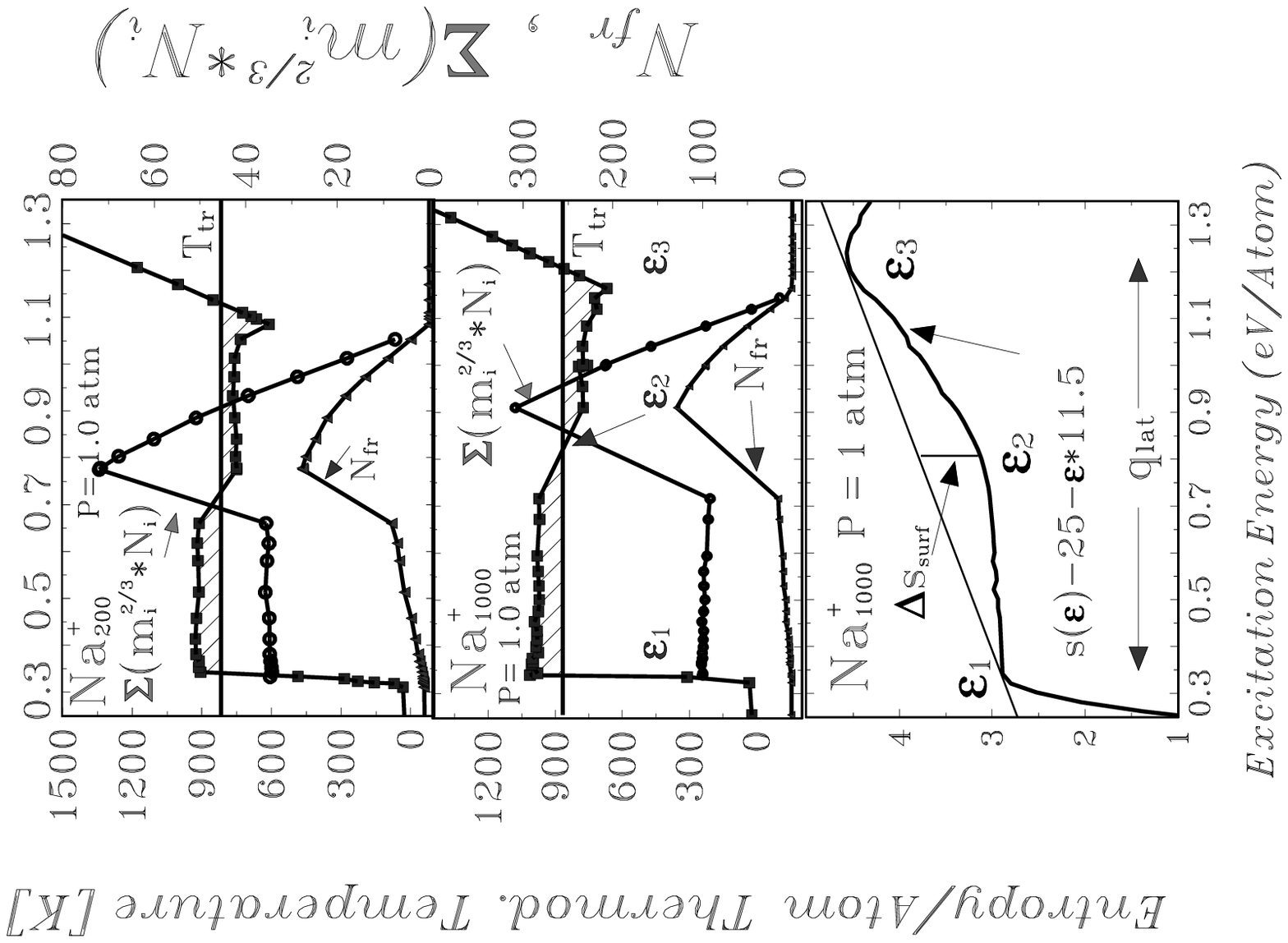}
\end{minipage}\rule{0.5cm}{0mm}\begin{minipage}[t]{5.5cm}
\includegraphics*[bb = 28 144 488 591, angle=-90, width=6cm,  
clip=true]{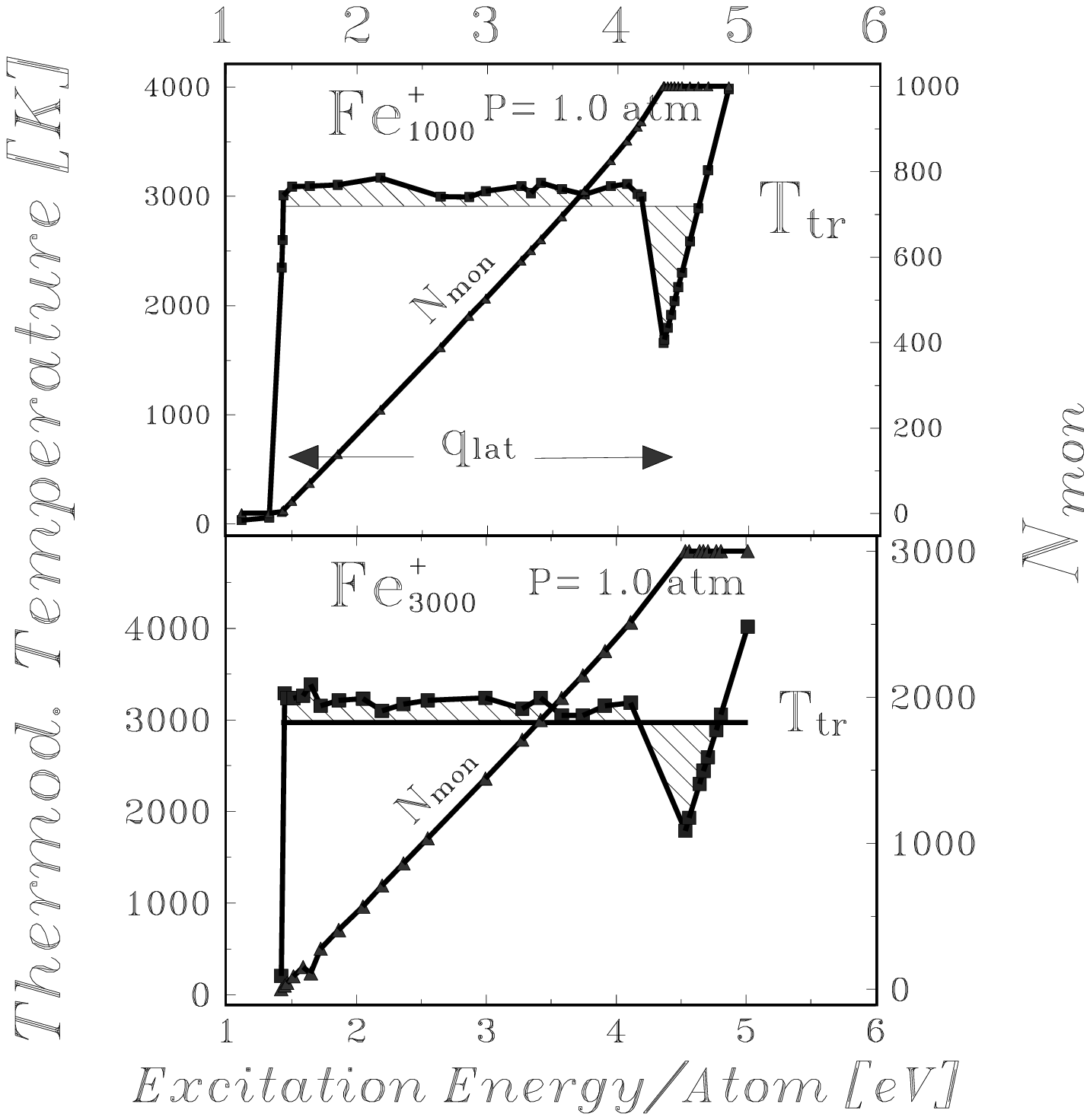} 
{\scriptsize Fig.2:~ microcanonical caloric curve $T_P(E/N= e)=(\partial
S/\partial E)^{-1}|_P$ at const. pressure (full square points), number of
monomers}
\end{minipage}\\
\begin{minipage}[h]{12.5cm}
{\scriptsize Fig.1:~microcanonical caloric curve
$T_P(E/N= e,V(E,P,N),N)=(\partial
S/\partial E)^{-1}|_P$ at constant pressure (full square
points), number of fragments $N_{fr}$ with $m_i\ge 2$ atoms, and the
effective number of surface atoms $N_{eff}^{2/3}=\sum m_i^{2/3} N_i=$
total surface area divided by $4\pi r_{ws}^2$. In the lower panel
$s( e)=\int_0^{ e}\beta( e')d e'$
is shown.  In order to make the intruder between $ e_1$ and
$ e_3$ visible, we subtracted the linear function
$25+11.5 e$.}
\end{minipage}\\~\\ 
{\em 2.4)~~ Geometric vs. entropic classification 
phases and phase-boundaries}\\ Our approach to phase-transitions of
first order is complementary to the conventional approach where the
separation of the system into two homogeneous phases by a --- in
general --- geometrical interface is investigated,
e.g.\cite{panagiotopoulos87}. The problems due to the large
fluctuations of this interface are numerous and severe, see e.g. the
discussion on fluid interfaces by Evans \cite{evans92}. These
fluctuation are of course crucial for the interfacial entropy and
consequently for the surface tension also.  

The two main differences of our approach are that a) we do not start with the
{\em geometry} (planar or spherical) of the interface, nor do we demand the
interface to compact. We focus our attention
to the {\em entropy} of the phase separation.  This turns out to be much
simpler than the geometric approach.  b) The microcanonical ensemble allows for
large scale spatial inhomogeneities, whereas the canonical ensemble {\em
suppresses} spatial inhomogeneities like phase-separations exponentially. Any
interface needs an additional free energy. \{In the case of a phase transition
of first order the suppression is $\propto exp(-\sigma N^{2/3}/T_{tr})$, where
$\sigma$ is the surface tension parameter ($\sigma =4\pi r_{ws}^2\gamma$) and
$r_{ws}$ is the radius of the Wigner-Seitz cell.

Our characterization of phase transitions is purely thermodynamically. We have
not yet defined what a phase is.  Much effort is spent by Ruelle to define pure
phases as those for which in the thermodynamic limit observables survive
increasing coarse-graining, for which space-averaged quantities do not
fluctuate, c.f.  chapter 6.5 in Ruelle's book \cite{ruelle69}. Of course, this
definition works in the thermodynamic limit only. It does not address finite
systems. For a finite system it is not possible to decide if a single
configuration corresponds to a pure phase or not. The situation is analogous to
the definition of the temperature, see above. Again to be a pure phase is a
feature of the whole ensemble not of a single phase-space point
(configuration).  We offer a {\em statistical} definition of a pure phase: A
configuration belongs to the ensemble of pure phases --- including its
fluctuations --- at concave points of $S(E)$ with $\partial^2 S/\partial E^2 <
0$.\\~\\ 
{\em 2.5)~~ The microcanonical constant pressure ensemble}\\ 
The micro-canonical ensemble with given pressure \{E,P,N\} must be
distinguished from the (in spirit) similar constant pressure ensemble \{H,P,N\}
introduced by Andersen \cite{andersen80,brown84,procacci94} where a
molecular-dynamic calculation with the hypothetical Hamiltonian \cite{brown84}
\begin{equation}
H(\{r_i,p_i\},V,\dot{V})=\frac{V^{2/3}}{2}\sum_{i=1}^{N}m\vecbm{$\dot{r_i}$}^2 + \sum_{i=1}^N\sum_{j>1}^N\Phi(r_{ij}V^{1/3})+
\frac{M}{2}\dot{V}^2+P_EV
\end{equation}
is suggested. Here $V$ is the volume of the system, taken as an
additional explicit degree of freedom, $\{r_i,p_i\}$ are the
coordinates and momenta of the atoms scaled with the factor $V^{1/3}$,
$\vecbm{$\dot{r_i}$},\dot{V}$ the corresp. velocities,
$\Phi(r_{ij})$ is the intra-atomic two-body potential, and $M$ is a
hypothetical mass for the volume degree of freedom. $P_E$ is the
given pressure. The total enthalpy $H$, atoms plus $V$-degree
of freedom, is conserved, not the total energy $E$ of the atoms alone. 

This is very different to our micro-canonical approach with given
$E$, $V(E,P)$ ,$N$ where the energy $E$ of the atoms is conserved and the
pressure is the correct thermodynamic pressure ($P(E,V)=T(E,V)\partial
S/\partial V|_E$). At each energy the volume $V(E,P)$ is simultaneously chosen
for all members of the ensemble by the condition that $T(E,V)\partial
S(E,V)/\partial V|_E$ of the whole ensemble is the correct pressure. In this
case there is a unique correlation between the energy $E$ and the volume $V$
which does not fluctuate within the ensemble even though the pressure is
specified. At the given energy this is still the \{E,V(E,P,N),N\} ensemble.\\

\section{The liquid-gas transition of sodium, potassium, and iron} 
The microscopic simulation of the liquid--gas transition in metals is
especially difficult.  Due to the delocalisation of the conductance electrons
metals are not bound by two-body forces but experience long-range many-body
interactions. Moreover, at the liquid--gas transition the binding changes from
metallic to covalent binding. This fact is a main obstacle for the conventional
treatment by molecular dynamics \cite{allen85}. 

In the macro-micro approach we do not follow each atom like in molecular
dynamics, the basic particles are the fragments. Their ground-state binding
energies are taken from experiments. The fragments are spherical and have
translational, rotational, and intrinsic degrees of freedom.  The internal
degrees of freedom of the fragments are simulated as pieces of bulk matter. The
internal density of states, resp. the internal entropy of the fragments is
taken as the specific bulk entropy $s( e)$ at excitation energies
$ e\le  e_{max}= e_{boil}$, which can be determined
from the experimentally known specific heat of the solid/liquid bulk matter
\cite{gross141}. $ e_{boil}$ is the specific energy where the boiling
of bulk matter starts. Details are discussed in
\cite{gross157,gross153}. Then the metallic binding poses no
difficulty for us and the metal --- nonmetal transitions is in our
approach controlled by the increasing fragmentation of the
system. This leads to a decreasing mean coordination number when the
transition is approached from the liquid side while the distance to
the nearest neighbor keeps about the same. Exactly this behavior was
recently experimentally observed \cite{hensel95,ross96}. By using the
microcanonical ensemble we do not prespecify the intra-phase surface
and allow it to take any form. Also any fragmentation of the interface
is allowed. It is the entropy alone which determines the fluctuations
of the interface. Here we present the first microscopic calculation of
the surface tension in liquid sodium, potassium, and iron.

The decay of potassium is in all details similar to that of
sodium,fig.(1). Therefore we don't show here the corresponding
figures. The liquid--gas transition in iron is different from that of
the alkali metals: Due to the considerably larger surface energy
parameter $a_s$ in the liquid drop formula of the ground-state binding
energies of iron compared to alkali metals there is no
multi-fragmentation of iron clusters at $P=1$ atm. Iron clusters of
$N\le 3000$ atoms decay by multiple monomer evaporation c.f. fig.2.
\begin{center} 
\begin{minipage}[h]{8cm}
\begin{tabular} {|c|c|c|c|c|c|} \hline 
&$N_0$&$200$&$1000$&$3000$&\vecb{bulk}\\ 
\hline 
\hline  
&$T_{tr} \;[K]$&$816$&$866$&$948$&\vecbm{$1156$}\\ \cline{2-6} 
&$q_{lat} \;[eV]$&$0.791$&$0.871$&$0.91$&\vecbm{$0.923$}\\ \cline{2-6} 
{\bf Na}&$s_{boil}$&$11.25$&$11.67$&$11.2$&\vecbm{$9.267$}\\ \cline{2-6} 
&$\Delta s_{surf}$&$0.55$&$0.56$&$0.45$&\\ \cline{2-6} 
&$N_{eff}^{2/3}$&$39.94$&$98.53$&$186.6$&\vecbm{$\infty$}\\
\cline{2-6} 
&$\sigma/T_{tr}$&$2.75$&$5.68$&$7.07$&\vecbm{$7.41$}\\ 
\hline 
\hline
&$T_{tr} \;[K]$&$697$&$767$&$832$&\vecbm{$1033$}\\ \cline{2-6}
&$q_{lat} \;[eV]$&$0.62$&$0.7$&$0.73$&\vecbm{$0.80$}\\ \cline{2-6}
{\bf K}&$s_{boil}$&$10.35$&$10.59$&$10.15$&\vecbm{$8.99$}\\ \cline{2-6}
&$\Delta s_{surf}$&$0.65$&$0.65$&$0.38$&\\ \cline{2-6}
&$N_{eff}^{2/3}$&$32.52$&$92.01$&$187$&\vecbm{$\infty$}\\ \cline{2-6}
&$\sigma/T_{tr}$&$3.99$&$7.06$&$6.06$&\vecbm{$7.31$}\\
\hline
\hline 
&$T_{tr} \;[K]$&$2600$&$2910$&$2971$&\vecbm{$3158$}\\ \cline{2-6} 
&$q_{lat} \;[eV]$&$2.77$&$3.18$&$3.34$&\vecbm{$3.55$}\\ \cline{2-6} 
{\bf Fe}&$s_{boil}$&$12.38$&$12.68$&$13.1$&\vecbm{$13.04$}\\ \cline{2-6} 
&$\Delta s_{surf}$&$0.75$&$0.58$&$0.77$&\\ \cline{2-6} 
&$N_{eff}^{2/3}$&$22.29$&$65.40$&$142.12$&\vecbm{$\infty$}\\
\cline{2-6} 
&$\sigma/T_{tr}$&$6.73$&$8.87$&$16.25$&\vecbm{$17.49$}\\ 
\hline 
\end{tabular}
\end{minipage}
\end{center}
{\bf TABLE 1:}{\scriptsize
Parameters of the liquid--gas transition at constant pressure 
of $1$ atm. in a microcanonical system of $N_0$ interacting atoms and
in the bulk. $s_{boil}=q_{lat}/T_{tr}$, it is interesting that the
value of $s_{boil}$ for all three systems and at all sizes is near to
$s_{boil}=10$ as proposed by the empirical Trouton's rule
\protect\cite{reif65}, $N_{eff}^{2/3}=\sum{m_i^{2/3} N_i}$, $N_i=$
multiplicity of fragments with $m_i$-atoms, and
$\sigma/T_{tr}=N_0\Delta s_{surf}/N_{eff}^{2/3}$.  The bulk values
$\sigma/T_{tr}$ are adjusted to the input values of $a_s$ taken for
the $T=0$ surface tension from ref.\protect\cite{brechignac95b} which
we used in the present calculation for the ground-state binding
energies of the fragments. $T_{tr}$ is $T_P$ not $T_V$. $T_V$ comes
even closer to the bulk value.Other inputs of the calculations are:
gound state liquid drop parameters of all possible clusters and the
bulk entropy/internal dof in the \underline{liquid phase} at $e\le
\varepsilon_1$ (fig.1).}

Table (1) gives a summary of all theoretical parameters for the
liquid-gas transition in clusters of $N_0=200-3000$ Na, K, and Fe atoms and
compared with their experimental bulk values. The transition-temperature
$T_{tr}$, the specific latent heat $q_{lat}$ and the entropy gain of an
evaporated atom $s_{boil}$ are approaching the experimental bulk values. 
$\Delta s_{surf}$ is the area under the back-bend of
$\beta( e)$. $N_0\Delta s_{surf}$ is the total entropy loss
due to the interfaces equal to 
$\sum 4\pi r_{ws}^2m_i^{2/3}N_i\gamma/T_{tr}=N_{eff}^{2/3}\sigma/T_{tr}$.
Of course, the transition temperature $T_{tr}$ and the latent
heat $q_{lat}$ of small clusters are smaller than the bulk values because
the average coordination number of an atom at the surface of a small
cluster is smaller than at a planar surface of the bulk. \\

\section{Conclusion}
We presented a formulation of statistical thermodynamics entirely from
the principles of mechanics. By carefully distinguishing conserved quantities
like the energy, or angular momentun from ensemble averaged quantities like
entropy, temperature or pressure this formulation allows an application to
small systems like nuclei or atomic clusters and to astrophysical systems.

The microcanonical statistics also gives a detailed insight into the nature of
phase transitions.  The liquid--gas transition in metals at normal pressure of
1 atm. is experimentally well explored. Therefore, it is a good test case for
our ideas and computational methods of microcanonical thermodynamics. By
allowing a system of $N=200-3000$ atoms to condense or fragment into an
arbitrary number of spherical fragment clusters and into an arbitrary number of
free atoms under a prescribed external pressure of 1 atm. we calculated by
microcanonical Monte Carlo methods \cite{gross153} the microcanonical caloric
curve $\beta( e)= \partial s/\partial e= <\partial/\partial  e>$. The anomaly
of $T( e)$, where $\partial T/\partial e \le 0$, signals the liquid--gas
phase-transition in the finite system. The characteristic parameters $T_{tr}$,
$q_{lat}$, and the surface tension $\sigma_{surf}$ are for $\sim 1000$ atoms
similar to the experimentally known values of the bulk liquid--gas transition.

This result is remarkable for several reasons:\\
a)~It proves that a phase transition of first order in a realistic continuous
system can very well be classified in small mesoscopic clusters {\em without
invoking the thermodynamic limit.} In fact, with the finite back-bending of
$T(E/N)$ the transition is easier recognizable than at $N\mra \infty$.\\
b)~Consistent with similar results for statistical toy models \cite{gross150}
$T_{tr},q_{lat}$,$\sigma_{surf}$ are for astonishing small systems close to
their bulk values. {\em The mechanism leading to phase-transition has nothing
to do with the thermodynamic limit.}\\
c)~Even for a realistic metallic system with its long-range many-body
interactions the {\em M}icrocanoncal {\em M}etropolis {\em M}onte {\em C}arlo
simulation method is able to describe the liquid--gas phase transitions quite
well.  This is possible because we do not use molecular dynamics with a
two-body Hamiltonian but use the experimental groundstate binding energies of
the fragment clusters which of course take care of the metallic bonding of
their constituents.\\
d)~The intra-phase surface entropy can be microscopically calculated.  This was
not possible up to now for metals. The surface tension per surface area can be
determined if the intra-phase {\em area} is known. For the surface entropy the
fluctuation and fragmentation of the surface are essential. At the liquid--gas
transition of sodium and potassium clusters of sizes as considered here at
normal pressure the intra-phase fluctuations are mainly due to strong
inhomogeneities and clusterization. This is consistent with recent experimental
evidence for the bulk \cite{hensel95,ross96}. This result is encouraging to
also compute the surface tension at higher pressure to see its vanishing
towards the critical point.\\ 
e)~The success of the {\em M}icrocanoncal {\em M}etropolis {\em M}onte
{\em C}arlo sampling method to reproduce for small clusters (within
$\le 20$\%) the known infinite matter values of the liquid--gas
transition is also a promising and necessary test of {\em MMMC} to
describe nuclear fragmentation \cite{gross95,gross153} and to be able
to get insight into the (critical) behavior of the nuclear matter
liquid--gas transition. This is important as it is obviously not
possible to compare the model with experimental data for nuclear
matter.\\
f)~The division of the macroscopic observables of the N-body system into
observables which can be determined at each phase-space {\em point}, in each
individual realization of the microcanonical ensemble, i.e.: globally conserved
quantities, energy, number of particles, charge, angular momentum,\\
{\em and} into thermodynamic observables which refer to the size of the 
ensemble,
the {\em volume} $e^S$ of the energy shell of the phase space, and which
cannot be determined at a single phase-space point, in a single event, like
entropy, temperature, pressure, chemical potential
is very essential. The concept of a pure phase and of a
phase-transition belongs to the second group.\\ 
g)~Last not least, there is a very fundamental difference in the microcanonical
and the canonical ensembles. The energy is the crucial parameter that controls
the fluctuations in their development from the liquid to the vapor and this is
respected only in the micro-ensemble.

I am grateful to Th. Klotz, V. Laliena, O. Fliegans, M. Madjet, and E. Votyakov
for many numerical computations and helpful discussions.


\end{document}